\documentclass{article}
\usepackage{arxiv}
\usepackage[utf8]{inputenc} 
\usepackage[T1]{fontenc}    
\usepackage{url}            
\usepackage{booktabs}       
\usepackage{amsfonts}       
\usepackage{amsmath}
\usepackage{nicefrac}       
\usepackage{microtype}      
\usepackage{graphicx}
\usepackage{float}
\usepackage{pgfplots}
\usepackage{caption}
\usepackage{tikz}
\pgfplotsset{compat=1.14}
\usetikzlibrary{plotmarks,shapes}
\usetikzlibrary{external}
\usepackage{hyperref}
\hypersetup{
    colorlinks=true,
    linkcolor=blue,
    filecolor=magenta,      
    urlcolor=cyan,
}

\title{Capacity increases obtained extending the transmission bandwidth in optical communication systems}

\author{
  Gabriel Saavedra \\
  \texttt{gasaavedra@udec.cl} \\
  Electrical Engineering Department\\
  Universidad de Concepcion
   \And
 Daniel Semrau \\
 \texttt{ daniel.semrau.15@ucl.ac.uk} \\
 Optical Networks Group\\
 University College London, London \\
 \And
 Polina Bayvel \\
  \texttt{ p.bayvel@ucl.ac.uk}\\
  Optical Networks Group\\
  University College London, London
}

\begin{document}
\maketitle

\begin{abstract}
    The potential benefits of extending the optical fibre transmission bandwidth are studied. Even in the presence of Kerr nonlinearity and inter-channel stimulated Raman scattering, increasing the usable optical fibre bandwidth appears to be the most promising solution to increase system throughput.
\end{abstract}

\keywords{Optical fibre communication \and Raman scaterring \and Fibre nonlinearity}

\section{Introduction}

Single-mode optical fibres (SMF) are the foundation of high-speed digital communication infrastructure, with a ubiquitous presence in today's interconnected society. They enable internet in our homes and are in submarine cables transferring data between continents. Since the introduction of SMF numerous developments have allowed the capacity of optical communication systems to increase over the years by some 6 orders of magnitude since the initial demonstrations in 1970s: the optical fibre amplifier has removed the need for the electronic repeaters and enabled the amplification of multiple, or multiplexed, signals with a single device; the coherent receiver enabled the use of different degrees of freedom of the optical field to transmit information; and information theory has developed coding schemes that allow an increase in the information carried per second and the correction of errors. Despite all of these, the trend observed during the first 40 years in optical communication systems, where the total throughput of the system was increased year after year, appears to be slowing down. The reason is the nonlinear nature of silica-based single-mode fibres that currently imposes a limit on the information they can carry as a communication channel, by introducing signal distortions that depend on the signal power and optical bandwidth. The limit imposed by fibre nonlinearities is sometimes referred to as the nonlinear Shannon limit or the capacity crunch \cite{Mitra2001,Essiambre2010}. Experimental demonstrations such as \cite{Cai1,Cai2,Ionescu} have used a variety of digital signal processing techniques to increase the total system throughput, however, the benefits of these techniques remain relatively limited, and even combined, they are unlikely to satisfy the capacity demands for future networks.

Potential solutions to further increase the information throughput in optical communications systems, and thus sustain the ever-increasing demand for data transmission, have been proposed, and they include: the use of an alternative optical fibre designs with an air core, free of fibre nonlinearities \cite{Richardson2016}; the use of new degrees of freedom to transmit information, namely space (space division multiplexing or SDM) \cite{RichardsonSDM}; and, finally the expansion of the transmission bandwidth in current SMF-based systems \cite{Winzer20years}. The latter approach is one capable of readily reusing existing fibre infrastructure and may well be the most viable option, considering the latest advances in optical subsystems \cite{NapoliMultiband}.

Here, we analyse the benefits from extending the transmission bandwidth as a solution to increase the throughput of optical communication systems based on single-mode silica-based optical fibres. As the signal bandwidth is increased to 10 THz and beyond, the nonlinear effect of inter-channel stimulated Raman scattering (ISRS), becomes increasingly important, as it causes power to be transferred from high to low frequency channels and thus changing the nonlinear interactions along the fibre channel. Previous attempts have tried to answer this question, in \cite{BayvelCapacity} a throughput of 800 Tb/s was estimated for a bandwidth of 50 THz using several simplifications for the calculations and not including the ISRS effect, and in \cite{SemrauOpEx} the system throughput was estimated in the presence of ISRS for a bandwidth up to 20 THz reporting continuous throughput increase by using larger bandwidths.  
Here we quantify the potentially achievable throughputs for bandwidths up to 33 THz in a fibre channel impaired by amplified spontaneous emission (ASE) noise, Kerr nonlinearity and ISRS, considering wavelength dependent attenuation and nonlinear coefficient, and nonlinear signal-noise interactions to complement previous analysis.

\section{Methodology}
\subsection{Model}
It has been shown that the upper bound on the information capacity of SMF, assuming an additive white Gaussian noise (AWGN) channel \cite{Kramer} is given by:
\begin{equation}
   C=Blog_2 (1+SNR),
   \label{Eq.Capacity}
\end{equation}

where B and SNR represent the signal bandwidth and signal-to-noise ratio, respectively. From Eq.~\eqref{Eq.Capacity} it appears evident that an increase in the signal bandwidth, at a constant SNR, will results in greater capacity. However, the nonlinear nature of SMF makes the problem of maintaining the SNR over extended bandwidths almost impossible. Several models have been proposed throughout the years to describe the nonlinear behaviour of optical fibres including, but not limited to \cite{Splett,Tang,Chen,Poggiolini2012,Johannisson,Mecozzi,Secondini,Dar,Carena,Golani,Serena,Ghazisaeidi,Carena2}. The Gaussian noise (GN) model \cite{Poggiolini2012}, and variations thereof, have enjoyed widespread popularity due to their reduced complexity and accuracy. The main assumption of the GN model is that nonlinear distortions can be modelled as AWGN process that results in a so-called nonlinear interference (NLI) noise power, which can be conveniently used in combination with Eq. (1) to estimate the capacity of a fibre channel. Recently, due to the renewed interest in broadband transmission systems, the GN model has been extended to included impairments due to ISRS \cite{SemrauOpEx,SemrauISRS,Roberts,Cantono1,Cantono2,Rabbani}.

In this work we apply the GN model, both in its original form \cite{Poggiolini2012} and the modified ISRS-version from \cite{SemrauOpEx} to calculate a NLI coefficient for every multiplexed channel over a variety of transmission bandwidths to quantify the potential benefits of using an extended transmission window including all modelled effects to date. 
Both models have been shown to accurately model the behaviour of the system for bandwidths up to 7.3 THz \cite{SaavedraGN} and 9 THz \cite{SaavedraISRS}. The SNR of the i-th channel was calculated using:
\begin{equation}
SNR_i=\frac{P_{(ch,i )}}{P_{(ASE,i)}+P_{(s-s,i)}+P_{(s-n,i)}},
\label{Eq.SNR}
\end{equation}
where $P_{(ch,i )}$ is the channel power, $P_{(ASE,i)}$ is the ASE noise within the channel bandwidth, $P_{(s-s,i)}$ and $P_(s-n,i)$ are the NLI noise power arising from signal-signal and signal-noise nonlinear interactions, respectively. Details on the calculation from Eq. \eqref{Eq.SNR} can be found in \cite{SemrauOpEx,Poggiolini2012}.  

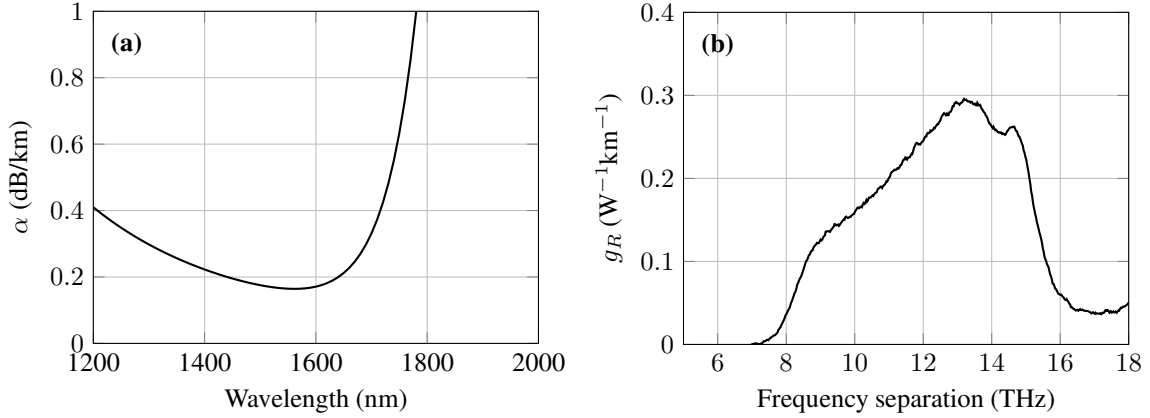
\begin{figure}[ht]
\centering
\begin{tikzpicture}
\begin{axis}
[width=7.5cm,height = 6cm,
xlabel={Wavelength (nm)},
xmin = 1200, xmax = 2000,
xtick={1200,1400,...,2000},
xticklabels = {1200,1400,1600,1800,2000},
ylabel={$\alpha$ (dB/km)},
ymin=0,ymax=1,
ytick = {0,0.2,...,2},
legend style={at={(0.25,0.95)},anchor=north},
xmajorgrids, ymajorgrids,
]
\node [text centered,align=center]  at (axis cs:1260,0.9) {\textbf{(a)}};
\addplot[color=black,thick] table [col sep=comma]{Data/Att_3.dat};
\end{axis};
\end{tikzpicture}
\begin{tikzpicture}
\centering
\begin{axis}
[width=7.5cm,height = 6cm,
ylabel={$g_R$ (W$^{-1}$km$^{-1}$)},
ymin = 0, ymax = 0.4,
xlabel={Frequency separation (THz)},
xmin=5,xmax=18,
xmajorgrids, ymajorgrids,
legend pos = north west,
]
\node [text centered,align=center]  at (axis cs:6,0.36) {\textbf{(b)}};
\addplot[color=black,thick] table [col sep=comma]{Data/RamanGain_Coeff.dat};
\end{axis};
\end{tikzpicture}
\caption{\textbf{(a)}Attenuation coefficient and \textbf{(b)} Raman gain coefficient for SMF.}
\label{Fig:gR}
\end{figure}

The signal power profiles in the presence of ISRS, were calculated using the following set of coupled differential equations \cite{Tariq93}:

\begin{equation}
\frac{dP_i}{dz} = -\sum_{j=i+1}^{M}\frac{\omega_j}{2\omega_i}\frac{g_R(\Omega)}{A_{eff}}P_jP_i + \sum_{j=1}^{i-1}\frac{g_R(\Omega)}{2A_{eff}}P_jP_i-\alpha_i P_i,
\label{eq:ISRS_eq}
\end{equation}  
where $i$ is the channel index being evaluated from a total of $M$ channels, $P_i$ is the optical power of channel $i$ , $\Omega$ is the frequency separation between channel $j$ and $i$, $\alpha_i$ is the attenuation coefficient of channel $i$, $\omega$ is the carrier frequency of the channel, and $g_R$ is the Raman gain coefficient. The channels are indexed such that the highest frequency channel has index $i = 1$. The attenuation and Raman gain coefficient as a function of wavelength and frequency separation, respectively, used for the calculations are show in Fig.~\ref{Fig:gR}.  

\subsection{Transmission system}

As a reference system a 2000~km SMF link was used. It is assumed that the system uses 80~km spans and the optical fibre is assumed to be low-loss with standard core area of 80~$\mu$m and attenuation coefficient of 0.165~dB/km at 1550~nm. 
Other liner impairments such as chromatic dispersion and polarisation mode dispersion were assumed to be compensated at the receiver using digital signal processing. Additionally, the dispersion parameter at 1480~nm was of 13.15~ps/(nm.km) with a slope of 0.092~ps/nm$^2$/km, and nonlinear coefficient ($\gamma$) of 1.2~1/(W.km). To calculate the NLI coefficient ($\eta$), a wavelength dependent nonlinear coefficient ($\gamma(\lambda)$) was used to include the variation of $\gamma$ across the studied spectrum. This was carried out by using a linear scaling relative to the increase of the mode field diameter (MFD) of the optical signal as a function of wavelength using 9.2 and 10.5~$\mu$m at 1310 and 1550~nm.     
The WDM channels are assumed to be shaped using an ideal root raised cosine filter, allowing for Nyquist channel spacing and modulation rate of 32~GBd. The optical power at the input of each fibre span was considered to be uniform across the spectrum. After each fibre span optical amplification is assumed to be carried out in parallel bands for different spectral regions of 5~THz after every span, performed by an ideal optical amplifier (noise figure of 3~dB) and maximum power of 27~dBm. This optical power was chosen due to the commercial availability of such powers in C- and L-band amplifiers. Other parameters used for the calculations are shown in Table~\ref{Table:parameters}.  

 \begin{table}[b]
\begin{center}
\begin{tabular}{ |c|c| } 
 \hline
 \textbf{Parameter}   				& \textbf{Value}  \\
  \hline
 Fibre Length         				& 80 (km) \\
 Centre Wavelength $(\lambda_c)$    & 1480 (nm)  \\ 
 Dispersion $(D_{\lambda_c)} $ 		& 13.15 (ps/nm/km)  \\ 
 Dispersion slope $(S_0)$			& 0.092 (ps/$\textnormal{nm}^2$/km) \\
 Nonlinear coefficient @ 1480 nm  ($\gamma$)   & 1.2  (1/W/km) \\
 Attenuation @ 1550 nm ($\alpha$)	& 0.165 (dB/km)\\
 Maximum Raman gain coefficient ($g_R$)     & 0.3  (1/W/km)\\
 Mode field diameter @ 1310 nm      & 9.2 ($\mu$m)\\
 Mode field diameter @ 1550 nm      & 10.5 ($\mu$m)\\ 				
 \hline
  \hline
  Symbol Rate 						& 32 [GBd]\\  
  Number of spans (N)				& 25 \\
  Maximum amplifier power			& 5 [W]    \\
  \hline
\end{tabular}
\end{center}
\caption{Transmission fibre parameters for ultra wideband transmission.}
\label{Table:parameters}
\end{table}

\section{Results}

\subsection{Linear impairments}
We first analysed the performance of a linear fibre channel, with Rayleigh backscattering and infra-red absorption considered to be the only causes of fibre attenuation across the spectrum\cite{Ohashi}. In this scenario, the signal-to-noise ratio (SNR) at the receiver is limited by the ASE noise from the optical amplification process and the output power available from the amplifiers. Any lumped optical amplifier introduces a total amount of ASE noise power given:
\begin{equation}
P_{(ASE,i)}=n_{sp} h\frac{\omega_i}{2\pi}(G-1)F_b,
\label{eq:ASE_EDFA}
\end{equation}
where $F_b$ is to the symbol rate of the channel centred at a frequency $\omega_i$, $n_{sp}$ is the spontaneous emission factor and $G$ is the amplifier gain at $\omega_i$. Eq.~\eqref{eq:ASE_EDFA} assumes that the ASE noise generated within a channel is be spectrally flat, however with the use of large optical bandwidths two terms from Eq.~\eqref{eq:ASE_EDFA} will significantly change for channels at different regions of the optical spectrum. Firstly, the central frequency ($\omega_0$) of the channels can change over 30~THz operating in the low-loss region of SMF, and secondly, the gain ($G$) required to amplify each channel is changed by the wavelength dependent attenuation coefficient. Using the attenuation coefficient from Fig.~\ref{Fig:gR}, the loss for each of the channels within the spectrum was calculated, and subsequently used to compute the amount of ASE generated by an ideal optical amplifier.
The SNR was then calculated as the ratio between received channel power and ASE noise in the channel bandwidth (with $P_{(s-s,i)}=P_{(s-n,i)}=0$ in Eq.~\eqref{Eq.SNR}) , where the received channel power was determined by the maximum amplifier output power.

\begin{figure}
\centering
\begin{tikzpicture}
\begin{axis}
[width=8cm,height = 6cm,
ylabel={SNR [dB]},
ymin = 0, ymax = 30,
ytick = {0,5,...,40},
yticklabels={0,5,10,15,20,25,30},
xlabel={Wavelength [nm]},
xmin=1230,xmax=1750,
xtick={1200,1300,...,1800},
xticklabels={1200,1300,1400,1500,1600,1700},
legend style={at={(0.2,0.9)},anchor=north},
xmajorgrids, ymajorgrids,
]
\addplot[color=blue,line width=0.25mm] table [col sep=comma]{Data/SNR_Linear.dat};
\end{axis};
\end{tikzpicture}
\caption{Signal to noise ratio in the presence of ASE only.}
\label{Fig:SNR_Lin}
\end{figure}
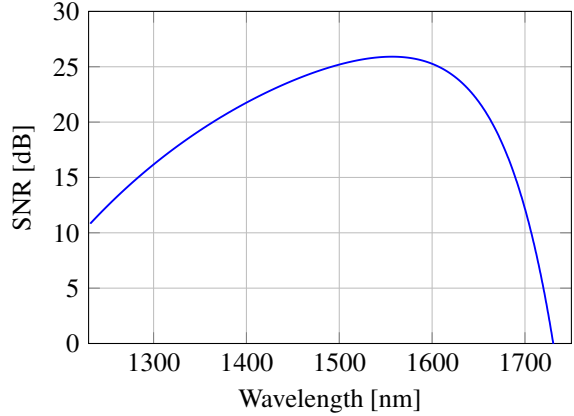

The received SNR for every channel placed between 1230 and 1750~nm can be seen in Fig.~\ref{Fig:SNR_Lin}. An increase in the SNR goes from short wavelengths towards the longer ones, following the attenuation profile of Rayleigh backscattering. A maximum SNR can be seen within the L-band at approximately 1580~nm, beyond this point SNR starts to decrease due to infra-red absorption. Increasing the transmission bandwidth beyond 1700~nm offers diminishing returns due to the rapid increase in the attenuation coefficient of optical fibres, and a SNR of 0~dB is found for wavelengths beyond 1745~nm for this particular configuration, rendering this region unusable.

\subsection{Kerr nonlinearity}

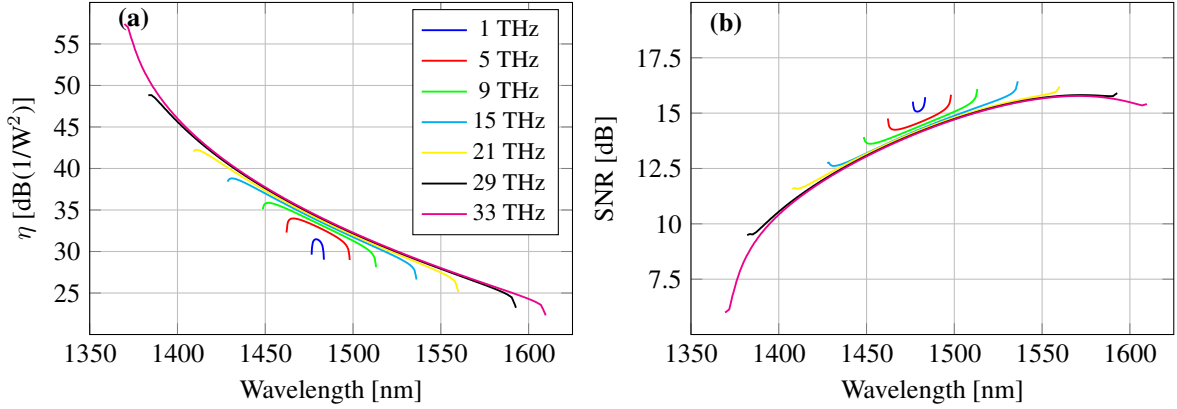
\begin{figure*}[t]
\centering
\begin{tikzpicture}
\begin{axis}
[width=8cm,height = 6cm,
ylabel={$\eta$ [dB(1/$\text{W}^2$)]},
ymin = 20, ymax = 60,
ytick = {25,30,...,55},
yticklabels = {25,30,35,40,45,50,55,60},
xlabel={Wavelength [nm]},
xmin=1350,xmax=1625,
xtick={1200,1250,...,1800},
xticklabels={1200,1250,1300,1350,1400,1450,1500,1550,1600,1650,1700},
legend style={at={(0.82,0.98)},anchor=north},
xmajorgrids, ymajorgrids,
]
\node [text centered,align=center]  at (axis cs:1375,58) {\textbf{(a)}};

\addplot[color=blue,line width=0.25mm] table [col sep=comma]{Data/ETA_1_THz.dat};
\addlegendentry{\small{1 THz}}
\addplot[color=red,line width=0.25mm] table [col sep=comma]{Data/ETA_5_THz.dat};
\addlegendentry{\small{5 THz}}
\addplot[color=green,line width=0.25mm] table [col sep=comma]{Data/ETA_9_THz.dat};
\addlegendentry{\small{9 THz}}
\addplot[color=cyan,line width=0.25mm] table [col sep=comma]{Data/ETA_15_THz.dat};
\addlegendentry{\small{15 THz}}
\addplot[color=yellow,line width=0.25mm] table [col sep=comma]{Data/ETA_21_THz.dat};
\addlegendentry{\small{21 THz}}
\addplot[color=black,line width=0.25mm] table [col sep=comma]{Data/ETA_29_THz.dat};
\addlegendentry{\small{29 THz}}
\addplot[color=magenta,line width=0.25mm] table [col sep=comma]{Data/ETA_33_THz.dat};
\addlegendentry{\small{33 THz}}
\end{axis};
\end{tikzpicture}
\begin{tikzpicture}
\begin{axis}
[width=8cm,height = 6cm,
ylabel={SNR [dB]},
ymin = 5, ymax =20,
ytick = {7.5,10,...,17.5},
yticklabels = {7.5,10,12.5,15,17.5},
xlabel={Wavelength [nm]},
xmin=1350,xmax=1625,
xtick={1200,1250,...,1800},
xticklabels={1200,1250,1300,1350,1400,1450,1500,1550,1600,1650,1700},
legend style={at={(0.8,0.95)},anchor=north},
xmajorgrids, ymajorgrids,
]
\node [text centered,align=center]  at (axis cs:1370,19) {\textbf{(b)}};
\addplot[color=blue,line width=0.25mm] table [col sep=comma]{Data/SNR_NL_1THz.dat};
\addplot[color=red,line width=0.25mm] table [col sep=comma]{Data/SNR_NL_5THz.dat};
\addplot[color=green,line width=0.25mm] table [col sep=comma]{Data/SNR_NL_9THz.dat};
\addplot[color=cyan,line width=0.25mm] table [col sep=comma]{Data/SNR_NL_15THz.dat};
\addplot[color=yellow,line width=0.25mm] table [col sep=comma]{Data/SNR_NL_21THz.dat};
\addplot[color=black,line width=0.25mm] table [col sep=comma]{Data/SNR_NL_29THz.dat};
\addplot[color=magenta,line width=0.25mm] table [col sep=comma]{Data/SNR_NL_33THz.dat};
\end{axis};
\end{tikzpicture}
\caption{\textbf{(a)}Nonlinear interference coefficient and \textbf{(b)} received SNR in the presence of Kerr nonlineartiry.}
\label{Fig:NL}
\end{figure*}

The NLI coefficient ($\eta$) was calculated using the GN-model after one span for bandwidths from 1 to 33~THz, using a central wavelength of 1480~nm. The maximum bandwidth was limited to avoid the zero dispersion wavelength of the optical fibre, region were the model has been shown to not accurately represent the behaviour of nonlinearity. NLI was assumed to add incoherently at every span, due to the use of large bandwidths. 
The NLI coefficients for the reference system are shown in Fig.~\ref{Fig:NL}~(a), and the corresponding received SNR at optimum power in Fig. \ref{Fig:NL}~(b). The optimum signal power was defined as the power that maximises the average SNR over all channels, and it was found to be -3 and -4~dBm for the studied bandwidths.  
For a fixed channel, as the signal bandwidth increased, the NLI continues to accumulate degrading the received SNR. For the bandwidths under study, the channels on the outer edge of the spectrum experienced smaller nonlinear distortions compared to the central ones. Compared to the linear case, nonlinearity clearly imposes a limit on the performance of the system by restricting the power into the fibre spans. The maximum SNR is reduced by  approximately 10~dB compared to the linear case due to nonlinear effects.
The tilt observed for the NLI coefficients,decreasing from shorter to longer wavelengths, is due to the dispersion slope of SMF and the growth of the MFD towards this region. For the largest studied bandwidth of 33 THz, the received SNR ranged from approximately 6~dB for the lower wavelengths up to 16 dB at 1580~nm.

Inter-channel stimulated Raman scattering becomes increasingly important as the transmission bandwidth grows. To study the impact of this effect, modifications to the GN model from\cite{SemrauOpEx} were used to estimate the system performance. The power transfer effect from ISRS modifies the signal power profile of every channel along the optical fibre, either reducing NLI if the channel is being depleted or increasing them if channel is being amplified with other cross-talk effects being negligible\cite{SaavedraISRS}. Additionally, the gain of the optical amplifiers after each span needs to be tailored to compensate the span loss and ISRS interactions for each channel. The overall ASE noise introduced by an optical amplifier after a fibre span including the ISRS effects is given by: 

\begin{equation}
P_{ASE,i}=n_{sp} h \frac{\omega_i}{2\pi}\left( \frac{G_{\alpha,i}}{G_{ISRS,i}}-1\right)F_{b},
\label{eq:Gain_ISRS_FullBW}
\end{equation} 
where $G_{\alpha,i}$ is the gain to compensate for the loss from fibre attenuation, and $G_{ISRS,i}$ is the gain to overcome the effect of ISRS. The term $G_{ISRS,i}$ is greater than 1 if the channel is amplified, thus gain smaller than fibre attenuation is required after the span. Conversely, the term $G_{ISRS,i}$ is smaller than 1 when the channel is depleted and a greater gain is required after the span. Therefore, ISRS modifies both the linear and nonlinear noise generation of the transmission system.

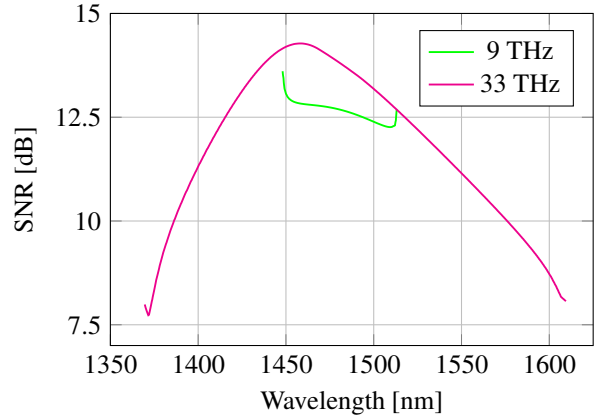
\begin{figure*}[b]
\centering
\begin{tikzpicture}
\begin{axis}
[width=8cm,height = 6cm,
ylabel={SNR [dB]},
ymin = 7, ymax =15,
ytick = {7.5,10,...,17.5},
yticklabels = {7.5,10,12.5,15,17.5},
xlabel={Wavelength [nm]},
xmin=1350,xmax=1625,
xtick={1200,1250,...,1800},
xticklabels={1200,1250,1300,1350,1400,1450,1500,1550,1600,1650,1700},
legend style={at={(0.8,0.95)},anchor=north},
xmajorgrids, ymajorgrids,
]
\node [text centered,align=center]  at (axis cs:1370,19) {\textbf{(b)}};
\addplot[color=green,line width=0.25mm] table []{Data/ISRS_9Thz.txt};
\addlegendentry{9 THz}
\addplot[color=magenta,line width=0.25mm] table [col sep=comma]{Data/33THz_optimum_1.dat};
\addlegendentry{33 THz}
\end{axis};
\end{tikzpicture}
\caption{Received SNR in the presence of ISRS for 9 and 33 THz.}
\label{Fig:NL_ISRS}
\end{figure*}

The received SNR as a function of channel wavelength at optimum power including ISRS is plotted in Fig.~\ref{Fig:NL_ISRS} for 9 and 33 THz. Due to ISRS the tilt observed in the received SNR at 9 THz is now in the opposite direction compared to the result from Fig. \ref{Fig:NL}~(b). At 33 THz an increase in the received SNR was observed for the lower wavelength region of the spectrum (between 1370 and 1450~nm), compared to Fig. \ref{Fig:NL}~(b). A maximum increase of 0.9 dB was found at 1450 nm. This is due to the smaller generation of NLI noise when the channel experiences depletion due to ISRS. Wavelengths longer than 1450 nm experienced a decrease in the received SNR due to ISRS. On average it was found that the received SNR, across the entire 33~THz spectrum, was reduced by 2.5~dB due to ISRS, compared to the study when ISRS was neglected. 

\subsection{Benefits of increasing the transmission bandwidth}

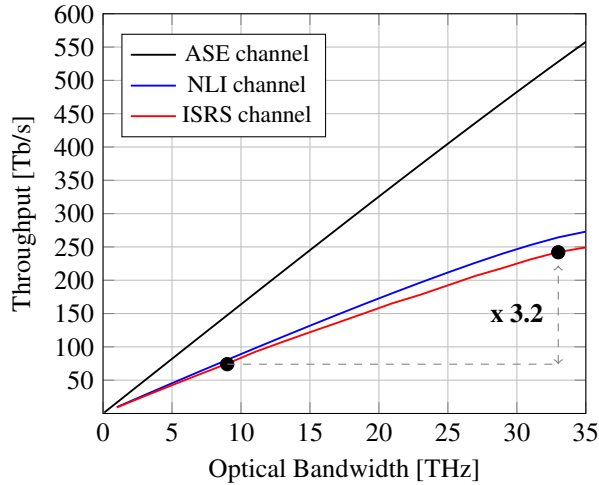
\begin{figure}
\centering
\begin{tikzpicture}
\begin{axis}
[width=8cm,
xlabel={Optical Bandwidth [THz]},
xmin = 0, xmax = 35,
xtick={0,5,...,40},
xticklabels = {0,5,10,15,20,25,30,35},
ylabel={Throughput [Tb/s]},
ymin=0,ymax=600,
ytick = {50,100,150,...,800},
yticklabels = {50,100,150,200,250,300,350,400,450,500,550,600},
legend style={at={(0.25,0.95)},anchor=north},
xmajorgrids, ymajorgrids,
]
\addplot[color=black,line width=0.25mm] table [col sep=comma]{Data/Throughput_Lin.dat};
\addlegendentry{\small{ASE channel}}
\addplot[color=blue,line width=0.25mm] table [col sep=comma]{Data/Throughput_NL.dat};
\addlegendentry{\small{NLI channel}}
\addplot[color=red,line width=0.25mm] table [col sep=comma]{Data/Throughput_ISRS.dat};
\addlegendentry{\small{ISRS channel}}
\node[circle,draw=black, fill=black, inner sep=0pt,minimum size=5pt] (a) at (9,74) {};
\node[circle,draw=black, fill=black, inner sep=0pt,minimum size=5pt] (b) at (33,242) {};
\draw[gray, dashed] (9,74)--(33,74);
\draw[gray, dashed,<->] (33,74)--(33,222);
\node [text centered,align=center]  at (axis cs:30,150) {\textbf{x 3.2}};
\end{axis};
\end{tikzpicture}
\caption{Channel throughput as a function of optical bandwidth.}
\label{Fig:Throughput}
\end{figure}

Extending the usable optical bandwidth will increase the generation and strength of nonlinear effects. However, as a solution to increase the capacity of modern optical communication systems using an extended optical bandwidth can still be an attractive option. NLI noise has an approximately logarithmic relationship with the bandwidth of the transmitted signals \cite{Poggiolini2012}. In any logarithmic function, the fastest growing stage is found close to zero, implying that the closest frequency components contribute the most to the generation of nonlinear distortions in a channel of interest, while channels with a large frequency separation present only a small contribution to it. However, ISRS modifies the generation of NLI along the transmission fibres making the answer of how much the throughput can be increased by using a larger optical bandwidth not simple.    

The potential achievable throughput as a function of optical bandwidth is plotted in Fig.~\ref{Fig:Throughput}. Due to the way the linear and NLI noise is treated in the used models, the throughput was estimated assuming an additive white Gaussian noise channel, therefore the potential system capacity was calculated using the results presented herein together with Eq.~\eqref{Eq.Capacity}, and is plotted in Fig.~\ref{Fig:Throughput} as a function of the optical bandwidth. Including ASE noise and fibre nonlinearities (including ISRS) a throughput of 242.2 Tb/s over 33 THz was found, which corresponds to 3.2 times increase compared to a system employing state -of-the-art bandwidth of 9 THz. In this multiband transmission regime the main challenge will be the effective control of the additional nonlinear distortions introduced by the ISRS effect. Even with the dramatic change in the power profiles of transmitted channels due to ISRS, and the implications on the nonlinear distortions, using a larger optical bandwidth leads to an increased system capacity. It thus remains attractive solution to increase the information throughput due to the approximately logarithmic relationship between nonlinear effects and the bandwidth of the transmitted signals. This will be especially true if the increases in point-to-point systems can be translated into increased throughput within the more complex network-wide operating regime.     

\section{Conclusion}
The potential benefits of increasing transmission bandwidth were quantified using the GN model. For all the studied scenarios increasing the transmission bandwidth resulted beneficial allowing a larger throughput to be transmitted. The throughput was increased approximately 3.2 times by extending bandwidth from 9 to 33~THz.  

\bibliographystyle{ieeetr}
\bibliography{IEEEabrv,bibliography}

\vspace{-4mm}

\end{document}